# Two Qubits in the Dirac Representation

A. K. Rajagopal[*] and R. W. Rendell
Naval Research Laboratory, Washington DC 20375-5320

## ABSTRACT

A general two qubit system expressed in terms of the complete set of unit and fifteen traceless, Hermitian Dirac matrices, is shown to exhibit novel features of this system. The well-known physical interpretations associated with the relativistic Dirac equation involving the symmetry operations of time-reversal T, charge conjugation C, parity P, and their products are **reinterpreted** here by examining their action on the basic Bell states. The transformation properties of the Bell basis states under these symmetry operations also reveal that C is the only operator that does not mix the Bell states whereas all others do. In a similar fashion, expressing the various logic gates introduced in the subject of quantum computers in terms of the Dirac matrices shows for example, that the NOT gate is related to the product of time-reversal and parity operators.

## I.   INTRODUCTION

A system of two qubits A and B offer a canonical paradigm for a large number of physical phenomena. The most general physical features of this system have been studied in terms of its density matrix $\hat{\Pi}(A,B)$, (which is a Hermitian, traceclass, and positive semi-definite matrix) written in terms of the parameters of single qubits of A and B and their mutual correlations. The presence of the mutual correlations in $\hat{\Pi}(A,B)$ indicate the entanglement property of the A and B qubits. Formal results concerning the characterization of the two qubit system have appeared in the literature and a recent account of this may be found in an elegant paper of Englert and Metwally [1]. Recently, Rau [2] has constructed propagators for a wide variety of qubit interactions. Dorai et al [3] have experimentally demonstrated the usefulness of selective NMR pulses in a quantum computation. Arvind and Mukunda [4] have given a two-qubit algorithm for finding quantum entanglement. The purpose of the present paper is to employ the complete set of sixteen Dirac matrices with a view to delve into possible physical insight that may emerge by reinterpreting the existing physical picture that underlies the symmetries of the relativistic Dirac theory, namely the time-reversal, charge conjugation, and parity suitably reinterpreted for the two qubit system. We follow and adopt the notations given in the review article of Good [5] for this purpose. An elegant implication of the algorithm in [4] is found when expressed in the Dirac representation and the associated symmetry.

In section II, we give the mathematical preliminaries and the associated physical interpretation of the general 4x4 matrices needed to characterize the two qubit system given in [1 (see references therein)] by expressing them in terms of the Dirac matrix representation. We give in this section an account of the important symmetry operations of time reversal (T), charge conjugation (C), and parity (P) and their products introduced in the development of relativistic theory and **reinterpret** them by working out their actions on the two qubit system expressed in terms of the basic Bell states. This leads us to a discussion of the special density matrices given in [1] in our language in section III. In section IV, in order to elucidate the symmetry properties of T, C, P of the Dirac theory in the context of the two-qubit system, we examine the properties of the Bell basis states under the action of these symmetry operations introduced above. These reveal several interesting features of the two qubit system. We establish the relationship between the Dirac and the Bell representations from which we display the physical significance of the Dirac operators in the context of quantum density matrix of two qubit systems. In section V, we express some logic gates given for example in [3], in terms of the Dirac matrices and examine their properties under the symmetry operations to explore further the underlying physical picture that may emerge in this viewpoint. In this section, we also express the results of [4] in the Dirac representation. We end the paper with a final section VI, where we make a few concluding remarks.

---

[*] rajagopa@estd.nrl.navy.mil

## II. TWO QUBITS IN DIRAC REPRESENTATION - T, C, P SYMMETRY

Each qubit is associated with a set of four 2x2 unit and (traceless) Pauli matrices:

$$\hat{1} = \begin{pmatrix} 1 & 0 \\ 0 & 1 \end{pmatrix}; \hat{\sigma}_1 = \begin{pmatrix} 0 & 1 \\ 1 & 0 \end{pmatrix}; \hat{\sigma}_2 = \begin{pmatrix} 0 & -i \\ i & 0 \end{pmatrix}; \hat{\sigma}_3 = \begin{pmatrix} 1 & 0 \\ 0 & -1 \end{pmatrix}. \tag{1}$$

Two qubits are then associated with sixteen 4x4 matrices formed by direct products of the Pauli matrices given above. These are listed in [2]. We note that the Dirac matrices are also formed by considering the same direct products. It is also well-known [5] that the Dirac theory of positrons and electrons leads to sets of four Hermitian, traceless matrices $\hat{\gamma}_\mu, \mu = 1, 2, 3, 4$ satisfying the relations $\hat{\gamma}_\mu \hat{\gamma}_\nu + \hat{\gamma}_\nu \hat{\gamma}_\mu = 2\delta_{\mu\nu}$. With these four, and their combinations, sixteen 4x4 traceless matrices $\hat{\gamma}_A$ such that each of its square is a unit matrix form a complete set in this 4x4 space. There are many explicit representations of these but we choose here the representation which gives the Schrodinger-Pauli limit of the Dirac theory and connects to Cayley-Klein parameters [5]. We list below the equivalent sixteen two qubit matrices with these Dirac matrices:

$$\hat{I} = \hat{I}(A) \otimes \hat{I}(B)$$

$$\hat{\gamma}_1 = \hat{\sigma}_2(A) \otimes \hat{\sigma}_1(B); \quad \hat{\gamma}_2 = \hat{\sigma}_2(A) \otimes \hat{\sigma}_2(B); \quad \hat{\gamma}_3 = \hat{\sigma}_2(A) \otimes \hat{\sigma}_3(B);$$
$$\hat{\gamma}_4 = -\hat{\sigma}_3(A) \otimes \hat{I}(B);$$

$$-i\hat{\gamma}_2\hat{\gamma}_3 = \begin{pmatrix} \hat{I}(A) \otimes \hat{\sigma}_1(B) \\ \equiv \hat{\Sigma}_1 \end{pmatrix}; \quad -i\hat{\gamma}_3\hat{\gamma}_1 = \begin{pmatrix} \hat{I}(A) \otimes \hat{\sigma}_2(B) \\ \equiv \hat{\Sigma}_2 \end{pmatrix}; \quad -i\hat{\gamma}_1\hat{\gamma}_2 = \begin{pmatrix} \hat{I}(A) \otimes \hat{\sigma}_3(B) \\ \equiv \hat{\Sigma}_3 \end{pmatrix};$$

$$i\hat{\gamma}_1\hat{\gamma}_4 = \hat{\sigma}_1(A) \otimes \hat{\sigma}_1(B); \quad i\hat{\gamma}_2\hat{\gamma}_4 = \hat{\sigma}_1(A) \otimes \hat{\sigma}_2(B); \quad i\hat{\gamma}_3\hat{\gamma}_4 = \hat{\sigma}_1(A) \otimes \hat{\sigma}_3(B);$$

$$i\hat{\gamma}_1\hat{\gamma}_5 = \hat{\sigma}_3(A) \otimes \hat{\sigma}_1(B); \quad i\hat{\gamma}_2\hat{\gamma}_5 = \hat{\sigma}_3(A) \otimes \hat{\sigma}_2(B); \quad i\hat{\gamma}_3\hat{\gamma}_5 = \hat{\sigma}_3(A) \otimes \hat{\sigma}_3(B);$$
$$i\hat{\gamma}_4\hat{\gamma}_5 = \hat{\sigma}_2(A) \otimes \hat{I}(B);$$

$$\hat{\gamma}_5 = \hat{\sigma}_1(A) \otimes \hat{I}(B). \tag{2}$$

We have arranged the above set to bring out the following known correspondence with the Dirac theory:
the first row is the 4x4 unit matrix;
the second row of four matrices form a four-vector;
the third row of six matrices form a second rank antisymmetric tensor;
the fourth row of four matrices form a four-pseudo-vector; and
the fifth row is a pseudo-scalar.
Furthermore, by considering the symmetry properties of the relativistic Dirac equation, the following important symmetries were found, which we **reinterpret** for the two qubit system here as follows:

$$\hat{T} \equiv \text{time reversal operator} = \hat{\gamma}_5\hat{\gamma}_4 = i\hat{\sigma}_2(A) \otimes \hat{I}(B);$$
$$\hat{C} \equiv \text{ch} \arg e \text{ conjugation} = -i\hat{\gamma}_2 = -i\hat{\sigma}_2(A) \otimes \hat{\sigma}_2(B); \text{ and} \tag{3a}$$
$$\hat{P} \equiv \text{parity operator} = i\hat{\gamma}_4 = -i\hat{\sigma}_3(A) \otimes \hat{I}(B);$$

Various products of these symmetries are also of interest and they are

$$\hat{T}\hat{C} = -\hat{\Sigma}_2; \quad \hat{C}\hat{P} = \hat{\gamma}_2\hat{\gamma}_4; \quad \hat{P}\hat{T} = -i\hat{\gamma}_5; \text{ and}$$
$$\hat{T}\hat{C}\hat{P} = \hat{\gamma}_2\hat{\gamma}_5. \tag{3b}$$

In the next section we will use these to construct the density matrices and their transformation properties.

## III. DENSITY MATRIX OF A GENERAL TWO QUBIT SYSTEM

We begin by representing the 2x2 density matrices of qubits A and B denoted respectively by $\hat{\rho}(A), \hat{\rho}(B)$ with unit trace as follows:

$$\hat{\rho}(A) = \frac{1}{2}\left(\hat{1}(A) + s_i(A)\hat{\sigma}_i(A)\right), \ \hat{\rho}(B) = \frac{1}{2}\left(\hat{1}(B) + s_j(B)\hat{\sigma}_j(B)\right) \quad (4)$$

In eq.(4), repeated indices denote the sum over the components, 1,2,3. The 3-vectors with real components in (4) parametrize the density matrices. The hat denotes matrix in the above. In general these density matrices represent mixed states when the magnitude of the vectors are less than unity and pure states when they are unity. In the two qubit system, these density matrices are represented by 4x4 matrices with unit trace denoted by $\hat{\Pi}(A), \hat{\Pi}(B)$ and is given by the direct products as shown:

$$\hat{\Pi}(A) = \frac{1}{2}\left(\hat{\rho}(A) \otimes \hat{1}(B)\right), \ \hat{\Pi}(B) = \frac{1}{2}\left(\hat{1}(A) \otimes \hat{\rho}(B)\right). \quad (5)$$

It should be pointed out that these density matrices represent mixed states in the two-qubit space even if their parent one-qubit density matrix represents a pure state. In fact, a simple calculation with $\left(\hat{\rho}^2(A) = \hat{\rho}(A)\right)$ in the 2-qubit space leads to $\left(\hat{\Pi}^2(A) = \frac{1}{2}\hat{\Pi}(A)\right)$. This clearly demonstrates the subtle features of the 2-qubit space.

The 4x4 density matrix formed by a direct product of $\hat{\rho}(A), \hat{\rho}(B)$ is useful and it is

$$\begin{aligned}\hat{\Pi}_d(A,B) &\equiv \hat{\rho}(A) \otimes \hat{\rho}(B) \\ &= \frac{1}{4}\begin{pmatrix} \hat{1}(A) \otimes \hat{1}(B) + s_i(A)\left(\hat{\sigma}_i(A) \otimes \hat{1}(B)\right) + \left(\hat{1}(A) \otimes \hat{\sigma}_j(B)\right)s_j(B) \\ + s_i(A)\left(\hat{\sigma}_i(A) \otimes \hat{\sigma}_j(B)\right)s_j(B) \end{pmatrix}\end{aligned} \quad (6)$$

This direct product density matrix is thus characterized basically by six real numbers plus nine others formed out of these as indicated by the last terms in eq.(6), which represent a simple form of the correlation between the two qubits in this density matrix. The most general density matrix describing two qubits however, has the following structure:

$$\hat{\Pi}(A,B) = \frac{1}{4}\begin{pmatrix} \hat{1}(A) \otimes \hat{1}(B) + s_i(A)\left(\hat{\sigma}_i(A) \otimes \hat{1}(B)\right) + \left(\hat{1}(A) \otimes \hat{\sigma}_j(B)\right)s_j(B) \\ + C_{ij}(A,B)\left(\hat{\sigma}_i(A) \otimes \hat{\sigma}_j(B)\right) \end{pmatrix} \quad (7)$$

Here $C_{ij}(A,B)$ are nine real numbers which are not factorized as in eq.(6), thus representing **correlation/entanglement** between the two qubits. We have chosen the structure of the most general two qubit density matrix in the above form so that the marginal density matrices are those given in eq.(4):

$$Tr_B \hat{\Pi}(A,B) = \hat{\rho}(A) \text{ and } Tr_A \hat{\Pi}(A,B) = \hat{\rho}(B). \tag{8}$$

If we subtract eq.(6) from eq.(7) we obtain an interesting expression exhibiting the correlation/entanglement between the two qubits explicity:

$$\hat{\Pi}(A,B) = \hat{\Pi}_d(A,B) + \frac{1}{4}\left(\left(C_{ij}(A,B) - s_i(A)s_j(B)\right)\left(\hat{\sigma}_i(A) \otimes \hat{\sigma}_j(B)\right)\right) \tag{9}$$

We now express eq.(7) in terms of the Dirac matrices by using eq.(2):

$$\hat{\Pi}(A,B) = \frac{1}{4}\begin{pmatrix} \hat{I} + s_1(A)\hat{\gamma}_5 + C_{2\mu}(A,B)\hat{\gamma}_\mu + iC_{3\mu}(A,B)\hat{\gamma}_\mu\hat{\gamma}_5 \\ + s_k(B)\hat{\Sigma}_k + iC_{1j}(A,B)\hat{\gamma}_j\hat{\gamma}_4 \end{pmatrix},$$

$$\text{where } C_{24}(A,B) = -s_3(A), \quad C_{34}(A,B) = s_2(A), \text{ and} \tag{10}$$

$$\hat{\Sigma}_k = -i\hat{\gamma}_i\hat{\gamma}_j \ (i,j,k \text{ cyclic}).$$

In this equation the repeated indices are summed with the notation introduced therein. In eq.(10), the second term is a pseudo-scalar, third term is a 4-vector, fourth is a pseudo-4-vector, and the the last two together form the second rank antisymmetric tensor. Note that our choice of the representation of the Dirac matrices explicitly displays the hermiticity of the density matrix.

It is worthwhile to display the one qubit systems given by eq.(5) in the two qubit space:

$$\hat{\Pi}(A) = \frac{1}{2}\left(\hat{\rho}(A) \otimes \hat{1}(B)\right) = \frac{1}{4}\left(\hat{I} + s_1(A)\hat{\gamma}_5 + s_2(A)i\hat{\gamma}_4\hat{\gamma}_5 - s_3(A)\hat{\gamma}_4\right)$$

$$= \frac{1}{4}\left(\hat{I} + s_1(A)i\hat{P}\hat{T} - s_2(A)i\hat{T} + s_3(A)i\hat{P}\right), \tag{11}$$

$$\hat{\Pi}(B) = \frac{1}{2}\left(\hat{1}(A) \otimes \hat{\rho}(B)\right) = \frac{1}{4}\left(\hat{I} + \vec{s}(B) \cdot \vec{\Sigma}\right).$$

The first expression for the density matrix of the A-qubit involves the pseudo-scalar, and the fourth components of the vector and pseudo-vector and in the second equivalent expression for it has an interesting structure involving the parity and time-reversal operators. The B-qubit, on the other hand, involves the components of the magnetization operator.
Similarly

$$\hat{\Pi}_d(A,B) = \frac{1}{4}\begin{pmatrix} \hat{I} + s_1(A)\hat{\gamma}_5 + is_2(A)\hat{\gamma}_4\hat{\gamma}_5 - s_3(A)\hat{\gamma}_4 + \vec{s}(B)\cdot\vec{\Sigma} \\ + is_1(A)\left(\vec{s}(B)\cdot\vec{\gamma}\right)\hat{\gamma}_4 + s_2(A)\left(\vec{s}(B)\cdot\vec{\gamma}\right) + is_3(A)\left(\vec{s}(B)\cdot\vec{\gamma}\right)\hat{\gamma}_5 \end{pmatrix}$$

$$\tag{12}$$

In the next section, we examine the action of the T, C, P operators on the basic Bell states.

## IV. T, C, P SYMMETRY PROPERTIES OF BELL BASIS STATES

The entanglement in the two qubit space is described by the complete set of orthonormal Bell-basis states which are given by

$$|\Psi_\pm\rangle = \frac{1}{\sqrt{2}}\left(|\uparrow A\rangle|\downarrow B\rangle \pm |\downarrow A\rangle|\uparrow B\rangle\right);$$
$$|\Phi_\pm\rangle = \frac{1}{\sqrt{2}}\left(|\uparrow A\rangle|\uparrow B\rangle \pm |\downarrow A\rangle|\downarrow B\rangle\right). \tag{13}$$

The action of the sixteen Dirac matrices on these states are found to be very useful in elucidating the viewpoint we are advocating here and are therefore given in Table I. These four states will now be used to rewrite the density matrices considered in the last section from which important conclusions concerning concepts of pure and mixed state entanglement will be deduced. By using the Bell operator, $\hat{B} = 2\sqrt{2}\left(|\Phi_+\rangle\langle\Phi_+| - |\Psi_-\rangle\langle\Psi_-|\right) = \sqrt{2}\left(\hat{\sigma}_1(A)\otimes\hat{\sigma}_1(B) + \hat{\sigma}_3(A)\otimes\hat{\sigma}_3(B)\right)$ and its square $\hat{B}^2 = 8\left(|\Phi_+\rangle\langle\Phi_+| + |\Psi_-\rangle\langle\Psi_-|\right) = 4\left(\hat{I}(A)\otimes\hat{I}(B) - \hat{\sigma}_2(A)\otimes\hat{\sigma}_2(B)\right)$, and expressing them in terms of the Dirac representation given in eq.(2), we obtain the following two pure state density matrices:

$$\hat{\Pi}_2(A,B) \equiv |\Psi_-\rangle\langle\Psi_-| = \frac{1}{4}\left(\hat{I} - i\hat{\gamma}_1\hat{\gamma}_4 - \hat{\gamma}_2 - i\hat{\gamma}_3\hat{\gamma}_5\right),$$
$$\hat{\Pi}_3(A,B) \equiv |\Phi_+\rangle\langle\Phi_+| = \frac{1}{4}\left(\hat{I} + i\hat{\gamma}_1\hat{\gamma}_4 - \hat{\gamma}_2 + i\hat{\gamma}_3\hat{\gamma}_5\right). \tag{14}$$

From the first line in Table I, we deduce the two other pure state density matrices

$$\hat{\Pi}_1(A,B) \equiv |\Psi_+\rangle\langle\Psi_+| = \frac{1}{4}\left(\hat{I} + i\hat{\gamma}_1\hat{\gamma}_4 + \hat{\gamma}_2 - i\hat{\gamma}_3\hat{\gamma}_5\right),$$
$$\hat{\Pi}_4(A,B) \equiv |\Phi_-\rangle\langle\Phi_-| = \frac{1}{4}\left(\hat{I} - i\hat{\gamma}_1\hat{\gamma}_4 + \hat{\gamma}_2 + i\hat{\gamma}_3\hat{\gamma}_5\right). \tag{15}$$

The sum of these four is the unit operator in the two qubit space, which is just the completeness property of the Bell basis set. Furthermore, twelve other off-diagonal operators of the form, $|\Psi_\pm\rangle\langle\Psi_m|, |\Psi_\pm\rangle\langle\Phi_\pm|, |\Phi_\pm\rangle\langle\Psi_\pm|, \text{ and } |\Phi_\pm\rangle\langle\Phi_m|$ are deduced from the transformation given in Table I starting from any one of the pure state density matrix given above. This set of sixteen density matrix operators based on the Bell states are just another equivalent expression for the sixteen operators of the Dirac representation. The relationship of the basic four $\hat{\gamma}$-matrices with the combinations of the outer products of Bell states is sufficient to establish the equivalent relationships between the two and glean from this a physical meaning that can be associated with the Dirac matrices in the two qubit context. We therefore list this here:

$$\hat{\gamma}_1 = i\big[|\Psi_+\rangle\langle\Psi_-| - |\Psi_-\rangle\langle\Psi_+| + |\Phi_+\rangle\langle\Phi_-| - |\Phi_-\rangle\langle\Phi_+|\big]$$

$$\hat{\gamma}_2 = \big[|\Psi_+\rangle\langle\Psi_+| - |\Psi_-\rangle\langle\Psi_-| - |\Phi_+\rangle\langle\Phi_+| + |\Phi_-\rangle\langle\Phi_-|\big] \quad (16)$$

$$\hat{\gamma}_3 = i\big[|\Psi_+\rangle\langle\Phi_+| - |\Psi_-\rangle\langle\Phi_-| - |\Phi_+\rangle\langle\Psi_+| + |\Phi_-\rangle\langle\Psi_-|\big]$$

$$\hat{\gamma}_4 = -\big[|\Psi_+\rangle\langle\Psi_-| + |\Psi_-\rangle\langle\Psi_+| + |\Phi_+\rangle\langle\Phi_-| + |\Phi_-\rangle\langle\Phi_+|\big]$$

From these it is interesting to observe that only $\hat{\gamma}_2$ has the diagonal Bell density matrices and all others represent "decoherent" parts of the density matrix in the Bell basis as they are all of the off-diagonal types. Moreover, another important feature is that the four Bell states are the eigen states of $\hat{\gamma}_2$.

The above four pure state density matrices represent entangled states because they have the following marginal density matrices of A and B qubits, which are maximally mixed chaotic states:

$$\hat{\rho}(A) = \frac{1}{2}\hat{I}(A) \text{ and } \hat{\rho}(B) = \frac{1}{2}\hat{I}(B). \quad (17)$$

We also observe that by comparing the four pure Bell-state density matrices in eqs.(14, 15) with eq.(10), they have besides the unit matrix, only the correlation terms of the form $C_{ii}, i = 1, 2, 3$, or in the Dirac representation, one part each of 3-vector, pseudo-3-vector, and 2nd rank antisymmetric tensor. This is a signature of the presence of an entangled state in a pure state density matrix in the two qubit case.

From Table I, we observe the following important relations:

$$\begin{aligned}
\hat{C}|\Psi_\pm\rangle &= \mp i|\Psi_\pm\rangle & \hat{C}|\Phi_\pm\rangle &= \pm i|\Phi_\pm\rangle \\
\hat{P}|\Psi_\pm\rangle &= -i|\Psi_m\rangle & \hat{P}|\Phi_\pm\rangle &= -i|\Phi_m\rangle \\
\hat{T}|\Psi_\pm\rangle &= \pm|\Phi_m\rangle & \hat{T}|\Phi_\pm\rangle &= \pm|\Psi_m\rangle \\
\hat{C}\hat{P}|\Psi_\pm\rangle &= \pm|\Psi_m\rangle & \hat{C}\hat{P}|\Phi_\pm\rangle &= \mp|\Phi_m\rangle \\
\hat{P}\hat{T}|\Psi_\pm\rangle &= \mp i|\Phi_\pm\rangle & \hat{P}\hat{T}|\Phi_\pm\rangle &= \mp i|\Psi_\pm\rangle & (18)\\
\hat{T}\hat{C}|\Psi_\pm\rangle &= i|\Phi_m\rangle & \hat{T}\hat{C}|\Phi_\pm\rangle &= -i|\Psi_m\rangle \\
\hat{C}\hat{P}\hat{T}|\Psi_\pm\rangle &= |\Phi_\pm\rangle & \hat{C}\hat{P}\hat{T}|\Phi_\pm\rangle &= -|\Psi_\pm\rangle
\end{aligned}$$

Note from this list, that $\hat{C}$ is the only operator that does not scramble the states whereas all others do. Thus the "charge conjugation" operator seems to have a unique property.

A third useful representation of the two qubit system is the total spin momentum representation, which here consists of singlet with zero spin, and three triplet states with spin one with components +1, -1, 0 of its projection along a pre-assigned z-axis. They are explicitly represented by

$$|s\rangle = |\Psi_-\rangle \, (\sin glet); \quad \begin{cases} |t_+\rangle = \frac{1}{\sqrt{2}}(|\Phi_+\rangle + |\Phi_-\rangle) \\ |t_0\rangle = |\Psi_+\rangle \\ |t_-\rangle = \frac{1}{\sqrt{2}}(|\Phi_+\rangle - |\Phi_-\rangle) \end{cases} (triplet) \qquad (19)$$

In this representation, the pure Bell state density matrices associated with the singlet and the triplet with zero projection remain as pure state density matrices and hence without decoherence. The other two pure Bell state density matrices have decoherent components in the two projection states of the triplet. In the Dirac representation, the unique operator $\hat{\gamma}_2$ is decoherent in the two projection states of the triplet:

$$\hat{\gamma}_2 = |t_0\rangle\langle t_0| - |s\rangle\langle s| - |t_+\rangle\langle t_+| - |t_-\rangle\langle t_-|. \qquad (20)$$

In the following section, we discuss the logic gates etc. in terms of the Dirac basis.

## V. DIRAC REPRESENTATION OF THE LOGIC GATES IN QUANTUM COMPUTATION

In this section we express several important logic gates in terms of the Dirac representation and provide another physical interpretation of these. We also include here a discussion of the Arvind-Mukunda algorithm in this frame work which reveals a surprising but simple interpretation of their result.

The NOT and Hadamard $(\hat{H})$ gates in the one-qubit representation are

$$NOT \equiv \hat{\sigma}_1 = (|\uparrow\rangle\langle\downarrow| + |\downarrow\rangle\langle\uparrow|);$$
$$\hat{H} \equiv \frac{1}{\sqrt{2}}(\hat{\sigma}_1 + \hat{\sigma}_3) = \frac{1}{\sqrt{2}}\left[(|\uparrow\rangle + |\downarrow\rangle)\langle\uparrow| + (|\uparrow\rangle - |\downarrow\rangle)\langle\downarrow|\right]. \qquad (21)$$

The two qubit gates are constructed as for example in [2,3], which we re-express in the Dirac representation.

$$CNOT \equiv XOR = \begin{pmatrix} 1 & 0 & 0 & 0 \\ 0 & 0 & 0 & 1 \\ 0 & 0 & 1 & 0 \\ 0 & 1 & 0 & 0 \end{pmatrix} = \frac{1}{2}(\hat{I} - i\hat{\gamma}_1\hat{\gamma}_2 + \hat{\gamma}_5 - i\hat{\gamma}_3\hat{\gamma}_4). \qquad (22)$$

Note that this contains the operator product of Parity and Time reversal operators.

$$NOT \equiv \begin{pmatrix} 0 & 0 & 1 & 0 \\ 0 & 0 & 0 & 1 \\ 1 & 0 & 0 & 0 \\ 0 & 1 & 0 & 0 \end{pmatrix} = \hat{\gamma}_5 \equiv i\hat{P}\hat{T}. \tag{23}$$

This NOT gate is just the product of Parity and Time reversal operators.

$$SWAP \equiv \begin{pmatrix} 1 & 0 & 0 & 0 \\ 0 & 0 & 1 & 0 \\ 0 & 1 & 0 & 0 \\ 0 & 0 & 0 & 1 \end{pmatrix} = \frac{1}{2}\left(\hat{I} + i\hat{\gamma}_1\hat{\gamma}_4 + \hat{\gamma}_2 + i\hat{\gamma}_3\hat{\gamma}_5\right). \tag{24}$$

This operator contains the charge conjugation operator. More interestingly, comparing with eqs.(13,14), we find that the SWAP gate is an algebraic sum of the pure Bell states:

$$SWAP = \hat{\Pi}_1 - \hat{\Pi}_2 + \hat{\Pi}_3 + \hat{\Pi}_4. \tag{25}$$

We now turn our attention to the algorithm suggested in [4] for distinguishing entangled from unentangled states, by classifying what they call even and odd density matrices. We first express the matrices given by them in our representation. This gives us an insight into the algorithm given in [4].
We have

$$\hat{\Pi}_{even} = \frac{1}{2}\begin{pmatrix} 1 & \pm 1 & 0 & 0 \\ \pm 1 & 1 & 0 & 0 \\ 0 & 0 & 0 & 0 \\ 0 & 0 & 0 & 0 \end{pmatrix} = \frac{1}{4}\left(\hat{I} - \hat{\gamma}_4 \mp i\hat{\gamma}_2\hat{\gamma}_3 \pm i\hat{\gamma}_1\hat{\gamma}_5\right). \tag{26}$$

$$\hat{\Pi}_{odd} = \frac{1}{2}\begin{pmatrix} 0 & 0 & 0 & 0 \\ 0 & 1 & \pm 1 & 0 \\ 0 & \pm 1 & 1 & 0 \\ 0 & 0 & 0 & 0 \end{pmatrix} = \frac{1}{4}\left(\hat{I} - i\hat{\gamma}_3\hat{\gamma}_5 \pm i\hat{\gamma}_1\hat{\gamma}_4 \pm \hat{\gamma}_2\right) \tag{27}$$

These are both pure state density matrices. We observe that under charge conjugation, $\hat{\Pi}_{odd}$ is invariant while $\hat{\Pi}_{even}$ is not (but $\hat{\Pi}_{even}$ is invariant under parity operation but not $\hat{\Pi}_{odd}$). Correspondingly the former is entangled because its marginal density matrix is a mixed state while the latter is not because its marginal density matrix remains a pure state. Moreover, by comparing with the pure Bell state density matrices (representing entangled pure states) in eqs.(14, 15), we find that $\hat{\Pi}_{odd}$ is $\hat{\Pi}_1$ if +sign is chosen, and $\hat{\Pi}_2$ if –sign is chosen. More generally, by examining eq.(7), in the light of eq.(27), the signature of entanglement apprears to be the presence of only $C_{ii}$ and all others absent, as was noted earlier in Sec.IV. On the other hand, $\hat{\Pi}_{even}$ is also a pure state expressed in terms of all the sixteen Bell state density matrices, diagonal and off-diagonal pieces. In fact, we find $\hat{\Pi}_{even} = |\Psi\rangle\langle\Psi|$, where
$|\Psi\rangle = |\Psi_+\rangle + |\Psi_-\rangle \pm |\Phi_+\rangle \pm |\Phi_-\rangle$.

We also observe that the eight even functions encoded by the unitary matrices given in [4], in the Dirac representation are

$$\hat{U}[0,4] = \hat{I} = -\hat{U}[4,0], \quad \hat{U}[2,2] = \pm\hat{\gamma}_4, \pm i\hat{\gamma}_1\hat{\gamma}_2, \pm i\hat{\gamma}_3\hat{\gamma}_5 \tag{28}$$

which are all invariant under the parity operation, P. We may also note that they are all separable in the sense that they can be respectively expressed as direct products as follows:

$$\hat{I}(A) \otimes \hat{I}(B), \hat{\sigma}_3(A) \otimes \hat{I}(B), \hat{I}(A) \otimes \hat{\sigma}_3(\mathbf{B}), \hat{\sigma}_3(A) \otimes \hat{\sigma}_3(\mathbf{B}). \tag{29}$$

The odd functions are more complicated and are not invariant under P. This is an alternate way to decide whether a given function is even or odd. Thus we have here provided an elegant insight into the algorithm given in [4].

## VI. CONCLUDING REMARKS

In order to gain clear insight into the physics of any underlying phenomena, the choice of a suitable representation is crucial. In the quantum theory of two-qubit systems, the Bell state and total spin representations have traditionally been employed with much success in understanding the subtle concepts of entanglement etc. In this paper, we have presented one more representation in terms of the Dirac Gamma matrices. We then reinterpret the underlying symmetries of the Dirac relativistic theory for the quantum two-qubit system. This has elucidated and shed light on some recent theoretical works on the subject as is evident from our Sec.V. We have also related this to the Bell states and the total angular momentum states to bring out the inter-relations among the various representations. Further implications of this representation, it is hoped, will be a fruitful avenue of investigation.

## ACKNOWLEDGEMENTS

Both the authors are supported in part by the Office Naval Research. We also thank Dr. Peter Reynolds of the Office Naval Research for supporting this research.

## REFERENCES

1. B. G. Englert and N. Metawally, e-print : quant-ph/9912089 v2.
2. A. R. P. Rau, Phys. Rev. **A61**, 032301-1 (2000).
3. K. Dorai, Arvind, and A. Kumar, Phys. Rev. **A61**, 042306 (2000).
4. Arvind and N. Mukunda, e-print : quant-ph/0006069.
5. R. H. Good, Jr., Rev. Mod. Phys. **27**, 187 (1955).

# TABLE I: Action of Dirac Matrices on the Bell states

The transformation under C, P, T, CP, PT, TC, TCP are shown in parenthesis.

$$\hat{\gamma}_1|\Psi_\pm\rangle = mi|\Psi_m\rangle \qquad\qquad \hat{\gamma}_1|\Phi_\pm\rangle = mi|\Phi_m\rangle$$

$$\hat{\gamma}_2|\Psi_\pm\rangle = \pm|\Psi_\pm\rangle \qquad (\hat{C} = -i\hat{\gamma}_2) \qquad \hat{\gamma}_2|\Phi_\pm\rangle = m|\Phi_\pm\rangle$$

$$\hat{\gamma}_3|\Psi_\pm\rangle = mi|\Phi_\pm\rangle \qquad\qquad \hat{\gamma}_3|\Phi_\pm\rangle = \pm i|\Psi_\pm\rangle$$

$$\hat{\gamma}_4|\Psi_\pm\rangle = -|\Psi_m\rangle \qquad (\hat{P} = i\hat{\gamma}_4) \qquad \hat{\gamma}_4|\Phi_\pm\rangle = -|\Phi_m\rangle$$

$$\hat{\gamma}_5|\Psi_\pm\rangle = \pm|\Phi_\pm\rangle \qquad (\hat{P}\hat{T} = -i\hat{\gamma}_5) \qquad \hat{\gamma}_5|\Phi_\pm\rangle = \pm|\Psi_\pm\rangle$$

$$\hat{\Sigma}_1|\Psi_\pm\rangle = |\Phi_\pm\rangle \qquad\qquad \hat{\Sigma}_1|\Phi_\pm\rangle = |\Psi_\pm\rangle$$

$$\hat{\Sigma}_2|\Psi_\pm\rangle = -i|\Phi_m\rangle \qquad (\hat{T}\hat{C} = -\hat{\Sigma}_2) \qquad \hat{\Sigma}_2|\Phi_\pm\rangle = i|\Psi_m\rangle$$

$$\hat{\Sigma}_3|\Psi_\pm\rangle = -|\Psi_m\rangle \qquad\qquad \hat{\Sigma}_3|\Phi_\pm\rangle = |\Phi_m\rangle$$

$$-i\hat{\gamma}_1\hat{\gamma}_4|\Psi_\pm\rangle = m|\Psi_\pm\rangle \qquad\qquad -i\hat{\gamma}_1\hat{\gamma}_4|\Phi_\pm\rangle = m|\Phi_\pm\rangle$$

$$-i\hat{\gamma}_2\hat{\gamma}_4|\Psi_\pm\rangle = mi|\Psi_m\rangle \quad (\hat{C}\hat{P} = \hat{\gamma}_2\hat{\gamma}_4) \quad -i\hat{\gamma}_2\hat{\gamma}_4|\Phi_\pm\rangle = \pm i|\Phi_m\rangle$$

$$-i\hat{\gamma}_3\hat{\gamma}_4|\Psi_\pm\rangle = m|\Phi_m\rangle \qquad\qquad -i\hat{\gamma}_3\hat{\gamma}_4|\Phi_\pm\rangle = \pm|\Psi_m\rangle$$

$$i\hat{\gamma}_1\hat{\gamma}_5|\Psi_\pm\rangle = |\Phi_m\rangle \qquad\qquad i\hat{\gamma}_1\hat{\gamma}_5|\Phi_\pm\rangle = |\Psi_m\rangle$$

$$i\hat{\gamma}_2\hat{\gamma}_5|\Psi_\pm\rangle = -i|\Phi_\pm\rangle \qquad (\hat{C}\hat{P}\hat{T} = -\hat{\gamma}_2\hat{\gamma}_5) \quad i\hat{\gamma}_2\hat{\gamma}_5|\Phi_\pm\rangle = i|\Psi_\pm\rangle$$

$$i\hat{\gamma}_3\hat{\gamma}_5|\Psi_\pm\rangle = -|\Psi_\pm\rangle \qquad\qquad i\hat{\gamma}_3\hat{\gamma}_5|\Phi_\pm\rangle = |\Phi_\pm\rangle$$

$$i\hat{\gamma}_4\hat{\gamma}_5|\Psi_\pm\rangle = mi|\Phi_m\rangle \qquad (\hat{T} = -\hat{\gamma}_4\hat{\gamma}_5) \qquad i\hat{\gamma}_4\hat{\gamma}_5|\Phi_\pm\rangle = mi|\Psi_m\rangle$$